\documentclass[aps,pre,twocolumn,showpacs,floatfix,superscriptaddress]{revtex4}
 \usepackage{graphicx}
\usepackage{epsfig}

\begin{document}

\title{Network Analysis of the State Space of Discrete Dynamical Systems} 

\author{Amer Shreim}
\affiliation{Complexity Science Group, Department of Physics and Astronomy,
University of Calgary, Calgary, Alberta, Canada}

\author{Peter Grassberger}
\affiliation{Complexity Science Group, Department of Physics and Astronomy,
University of Calgary, Calgary, Alberta, Canada}

\author{Walter Nadler}
\affiliation{Department of Physics, Michigan Technological University, Houghton, Michigan, USA}

\author{Bj{\"o}rn Samuelsson}
\affiliation{Department of Physics, Duke University, Durham, North Carolina, USA}

\author{Joshua E.S. Socolar}
\affiliation{Department of Physics, Duke University, Durham, North Carolina, USA}

\author{Maya Paczuski}
\affiliation{Complexity Science Group, Department of Physics and Astronomy,
University of Calgary, Calgary, Alberta, Canada}

\date{\today}

\begin{abstract}
  We study networks representing the dynamics of elementary 1-d
  cellular automata (CA) on finite lattices. We analyze scaling
  behaviors of both local and global network properties as a function of
  system size. The scaling of the largest node in-degree is obtained analytically
  for a variety of CA including rules 22, 54 and 110.
    We further define the
  \emph{path diversity} as a global network measure.
  The co-appearance of
  non-trivial scaling in both  hub size and  path diversity 
  separates
  simple dynamics from the more complex behaviors typically found in Wolfram's Class IV and some
  Class III CA.

\end{abstract}

\maketitle

As physical theories widen into the biological and social realms, the
problem of characterizing complex dynamical systems becomes
increasingly important.  Even elementary systems, such as cellular
automata (CA), originally proposed by von Neumann~\cite{VonNeuman},
often exhibit dynamical patterns that pose conceptual challenges and
 serve as test-beds for techniques to study more realistic
phenomena.  To date, attempts to classify the behavior of dynamical 
systems such as CA have been based on various definitions of 
complexity and have largely focused on patterns
generated in space and time (see, e.g.,
Ref.~\cite{Peter1986IntJPhys, Huberman, Badii,
  Bialek,Crutchfield}).  Here we take a different approach, focusing
on statistical properties of state space networks.

The trajectories of a discrete dynamical system form a directed
network in state space, wherein each node, representing a state, is
the source of a link that points to its dynamical successor~\cite{Wuensche}.  For deterministic systems, each node has a single
outgoing link (each out-degree is equal to 1). For irreversible systems,
states may have different numbers of pre-images and thus different
in-degrees.  In this paper we present analytical and numerical studies
of state space networks of various one-dimensional CA.  This
network perspective reveals previously unrecognized
 scaling behaviors and suggests
a new measure of an aspect of 
complexity that we term ``path diversity.''  Since
the CA we study
are known to produce a wide variety of different dynamical behaviors, 
our results are relevant for understanding
discrete dynamical systems in general.

We examine 1-d binary CA with nearest neighbor interactions and
periodic boundary conditions.  Wolfram put these CA into four
complexity classes~\cite{wolfram1984} based on the qualitative
appearance of spatio-temporal patterns produced from random initial
conditions for large lattice size $L$.  The four classes are: (I) the system almost
always evolves quickly to a unique fixed point; (II) it almost always
evolves quickly to one of many attractors with a small period; (III)
it generates seemingly random patterns with small-scale structures;
(IV) it shows a mixture of order and randomness with long
characteristic times.  One class IV CA, Rule 110 (defined below), has
been shown to emulate a universal Turing machine~\cite{Cook}.
One  shortcoming of this classification is that the border between
classes III and IV is ill-defined. In fact, the classification of some rules
(such as rule 54) is still disputed, and
 misclassifications can happen due to e.g. subtle and
 slow dynamics hidden 
beneath clear chaotic behavior. Examples include rule 18, where 
annihilating random walks are embedded in a random pattern with 
small-scale structure~\cite{Peter1983PRA}, and rule 22 which shows
random patterns with  slow but (highly) statistically  significant 
decrease of entropy with time and with $L$~\cite{Peter1986JStatPhys}.
Other classifications have been 
suggested, but a definitive criterion
for complex dynamics has not yet emerged~\cite{Cullik90,Gutowitz90,Dhar95}.

In the present Letter we report numerical and analytical results showing
that class IV and some class III CA exhibit highly
heterogeneous state space networks, unlike the networks corresponding to
the simple CA in class I and II. The heterogeneity is reflected in
local properties, including broad in-degree distributions and
finite-size scaling of the largest in-degree.  We show, however, that
this type of local heterogeneity can also occur in CA with simple
dynamics. On the other hand, a global measure, termed the
 \emph{path diversity}, 
by itself cannot distinguish simple from complex CA either.  However, it
tends to
show trivial behavior in simple CA where the local measures are non-trivial.
In fact, the  complex rules show non-trivial
scaling behavior in both the local measures and in the path diversity.
On the basis of these observations, we
 speculate that network heterogeneity at multiple levels may be 
a generic property of the state space of complex dynamical systems.

Let $R$ denote
the CA rule.  The binary value $s^{t+1}_i$ of site $i$ at time $t+1$ is 
set equal to $R(s^t_{i-1} s^t_i s^t_{i+1})$ evaluated at the previous 
time $t$.  We use periodic boundary conditions so that spatial indices 
are taken modulo $L$.  To
label the CA rules, we use Wolfram's scheme that identifies each rule
with the number $R(000) + 2R(001) + 2^2R(010) + 2^3R(011)+ 2^4R(100) +
2^5R(101) + 2^6R(110) + 2^7R(111)$.  Hence, for example, Rule 18 is the
one for which $R(100)=1$ and $R(001)=1$ and $R(s_1 s_2 s_3)=0$ for all
other triples.

\begin{figure}[htbp]
\begin{center}
\includegraphics*[width=7.7cm]{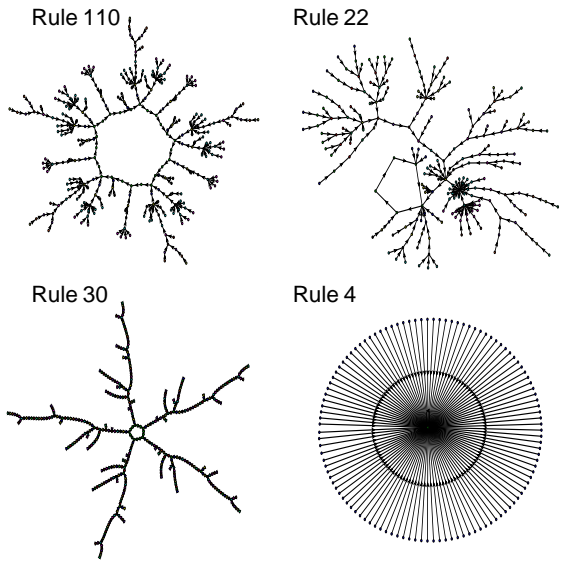}
\caption{\label{networks} One connected cluster out of each
state space network for different CA, plotted with the program `Pajek' 
(vlado.fmf.uni-lj.si/pub/networks/pajek/). All
arrows point from leaves towards the center. Sizes are $L = 10$ to 13.
For Rule 4, the network heterogeneity resides entirely in the size 
distribution of the different clusters making the state space network.}
\end{center}
\end{figure}

A binary CA of size $L$ has $N=2^L$ different states ${\bf S}^{(a)} =
(s_0^{(a)}\ldots s_{L-1}^{(a)}),\; a=0,\ldots,N-1$. These may be viewed
as the nodes of a directed network, where a link from ${\bf S}^{(a)}$
to ${\bf S}^{(b)}$ indicates that $R$ maps ${\bf S}^{(a)}$ to 
${\bf S}^{(b)}$. When such a link exists, we say that ${\bf S}^{(a)}$ 
is a pre-image of ${\bf S}^{(b)}$ and the number of pre-images of 
${\bf S}^{(b)}$ is its in-degree $k({\bf S}^{(b)})$. The network 
typically consists of disconnected clusters, each containing transient
states and a recurrent dynamical attractor. {\it Garden of Eden} (GoE)
states are transient states with zero in-degree. Pictures of some
state space clusters are shown in Fig.~\ref{networks}; for more, 
see~\cite{Wuensche}.

These state space networks can be analyzed by a variety of statistical 
measures -- e.g. their degree distribution, clustering coefficients, etc.
\cite{NewmanSIAMReview, AlbertBarabasiRevModPhys}. Many real-world 
networks (e.g. regulatory networks~\cite{Reka2005}, the world-wide 
web~\cite{BarabasiWWW}, and earthquakes~\cite{MayaNetworksEarthquakes})
differ markedly from random graphs
(where degree distributions are Poissonian and clustering is absent), 
displaying ``fat-tailed'' or even scale-free degree distributions.
In the following we show similar results also for state space networks. 



\begin{figure}[htbp]
\includegraphics*[width=2.7in, height=1.8in]{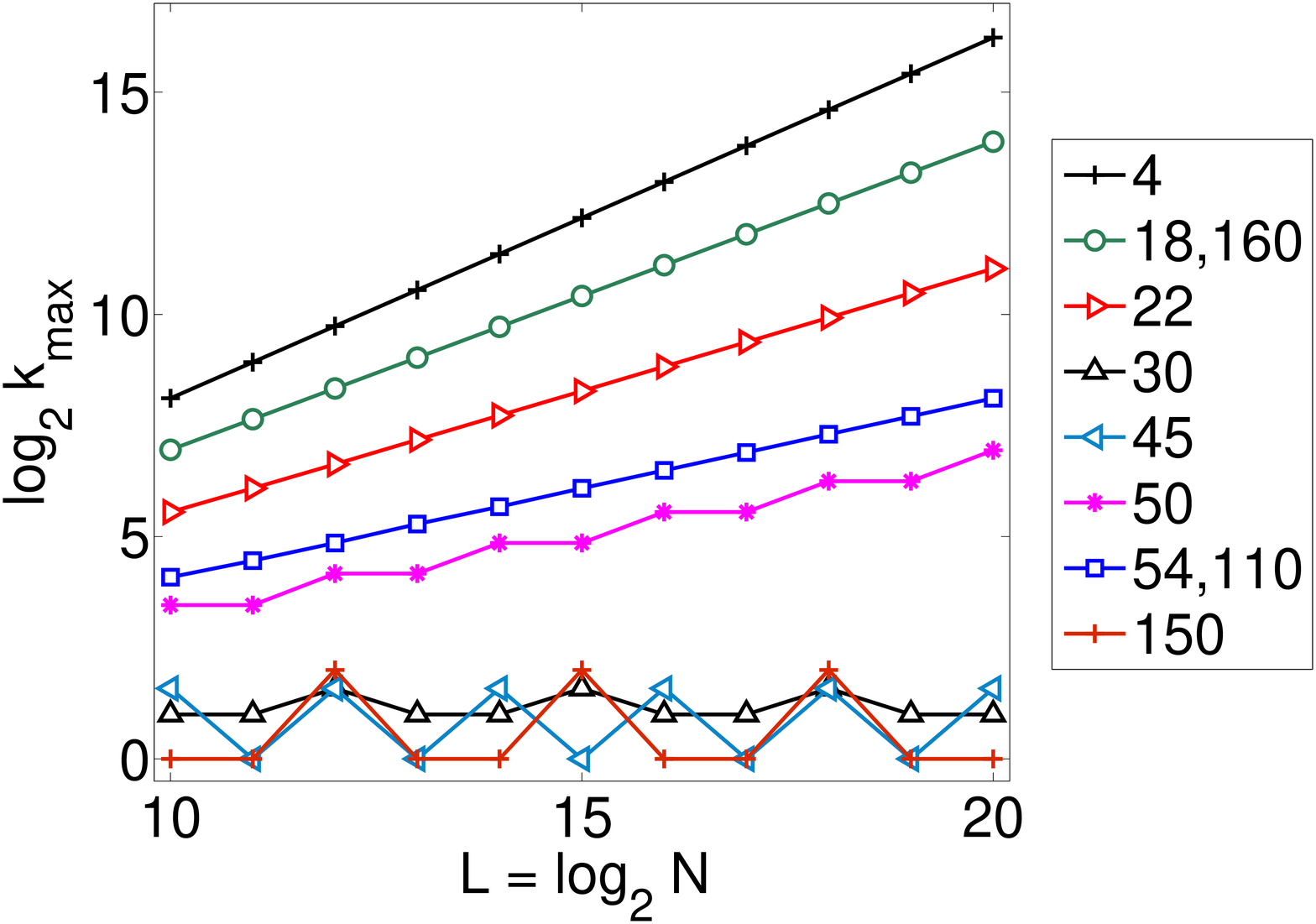}
\begin{center}
\caption{\label{largest_indegree} (color on-line) The largest in-degree $k_{max}$
  as a function of $N=2^L$. Except for class III rules 30, 45 and 150, all CA
  shown here exhibit clear scaling of the largest hub size $k_{max}$ with the total
  number of nodes in the network, $N$. Rule 160 is class I, 4 and 50 are class II,
  and 110 is class IV. Rules 18, 22, and 54 are between III and IV, as they 
  have large structures masked by small-scale chaos (18, 22) or 
  structures on intermediate scales (54). The analytic values of $\nu$ are
  $0.8114, 0.6942, 0.5515, 0.4057, 0.3471, 0.4057$ and $0.6942$ for
  rules 4, 18, 22, 50, 54, 110 and 160. These values are in perfect agreement
  with the numerical results.}
  \label{hubs}
\end{center}
\end{figure}

Fig.~\ref{largest_indegree} demonstrates that
several CA networks exhibit clean scaling for
a particular local property, the in-degree of the largest hub, $k_{max}
\sim N^{\nu}$. Scaling  sets in already for rather small lattices and
 holds also for the second, third, etc., largest hub (data not shown). 
The rules shown in Fig.~\ref{largest_indegree} were chosen to 
cover  the entire spectrum of known behaviors for
1-d elementary CA.

This scaling, including the value of $\nu$, can be derived exactly, 
provided one knows the structure of the hub state ${\cal H}$. The latter 
can be guessed easily for all CA shown in Fig.~\ref{largest_indegree}, 
while it is less obvious for others. For all elementary CA, the hub state is either
periodic or -- if the period does not match the lattice size -- periodic
with a few defects.

For rule 18, e.g., one finds numerically that ${\cal H}= (00\cdots 00)$ 
for all $L$. According to the definition of  CA 18 given above, all sequences
that do not contain 001 or 100 as substrings are pre-images of $\cal H$. 


\begin{figure}[htbp] 
\includegraphics*[width=2.8in, height=2.0cm]{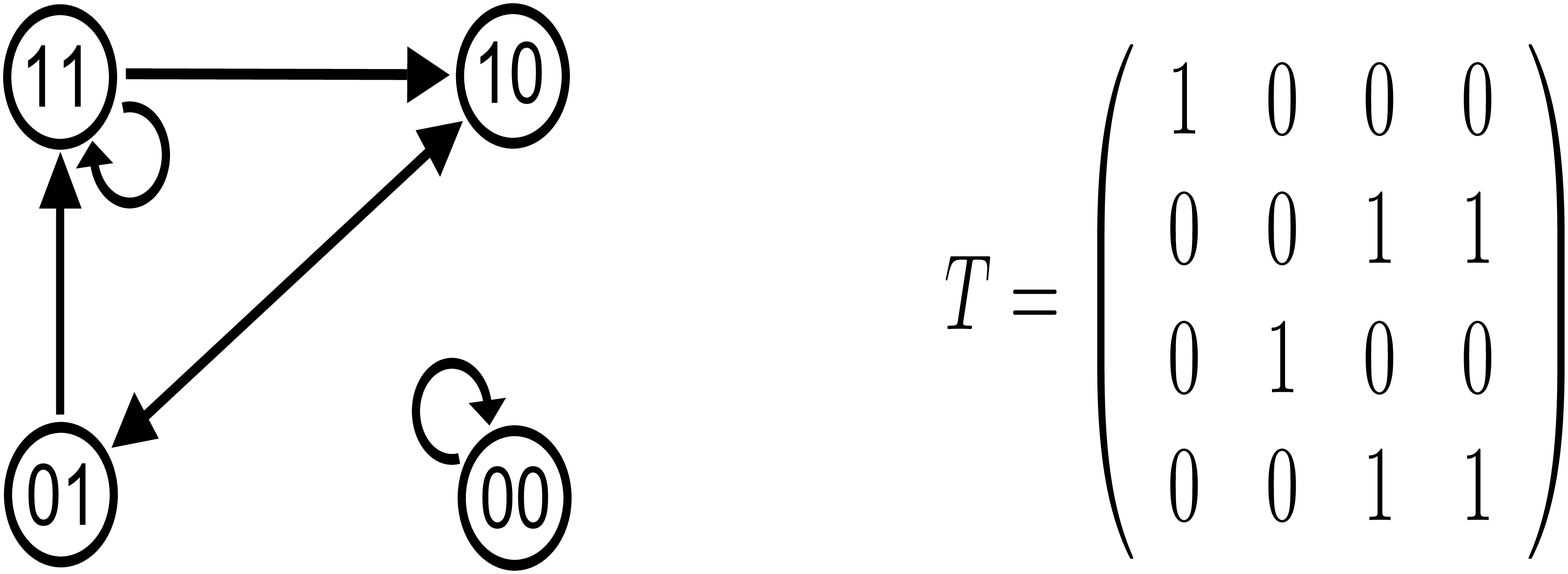}
  \caption{Walks on the
    graph shown on the left generate all pre-images of the hub state ${\cal H}$ 
    for rule 18. Each step corresponds to an element 
    of the matrix ${\bf T}$ shown on the right, with rows and columns 
    labeled in the order $00,01,10,11$.}
  \label{WalkRule18}
\end{figure}

The number of distinct strings of length $\ell$ without 
`001' or `100' is equal to the number of walks with $\ell-2$ steps on the 
graph shown in Fig.~\ref{WalkRule18}.
The number of periodic strings of length $L$ is then given by ${\rm Tr}{\bf T}^L 
\approx N^{\log_2\lambda_1}$, where ${\bf T}$ is the matrix 
shown on the right of Fig.~\ref{WalkRule18}, and $\lambda_1$ is its largest 
eigenvalue. This gives $k_{max} \sim N^{\nu}$ with $\nu = \log_2\lambda_1 
= 0.6942$, in perfect agreement with the numerical results.

Similar analytic 
arguments hold for all the other CA that exhibit scaling in 
Fig.~\ref{largest_indegree}. When the hub state contains both `0's and 
`1's, one has to introduce two transfer matrices ${\bf T}^{(0)}$ and 
${\bf T}^{(1)}$ where $T^{(s)}$ maps each pair $s_{i-1}^ts_i^t$ onto
the pair $s_i^ts_{i+1}^t$, iff $s_i^{t+1} \equiv R(s_{i-1}^ts_i^ts_{i+1}^t)
 = s$. The labeling of rows and columns is as in Fig.~\ref{WalkRule18}.
The in-degree of any state ${\bf S}$ is then
\begin{equation}
  k({\bf S})
      = {\rm Tr}({\bf T}^{(s_0)}\cdots{\bf T}^{(s_{L-1})})~.
\label{eq: k of S}
\end{equation}
The resulting exponents $\nu$ are cited in the caption for Fig.~\ref{hubs}.
The same scaling (with the same exponents) holds also for the second, third,
etc. largest hub.

\begin{figure}[htbp]
\begin{center}
\includegraphics*[width=2.7in, height=1.8in]{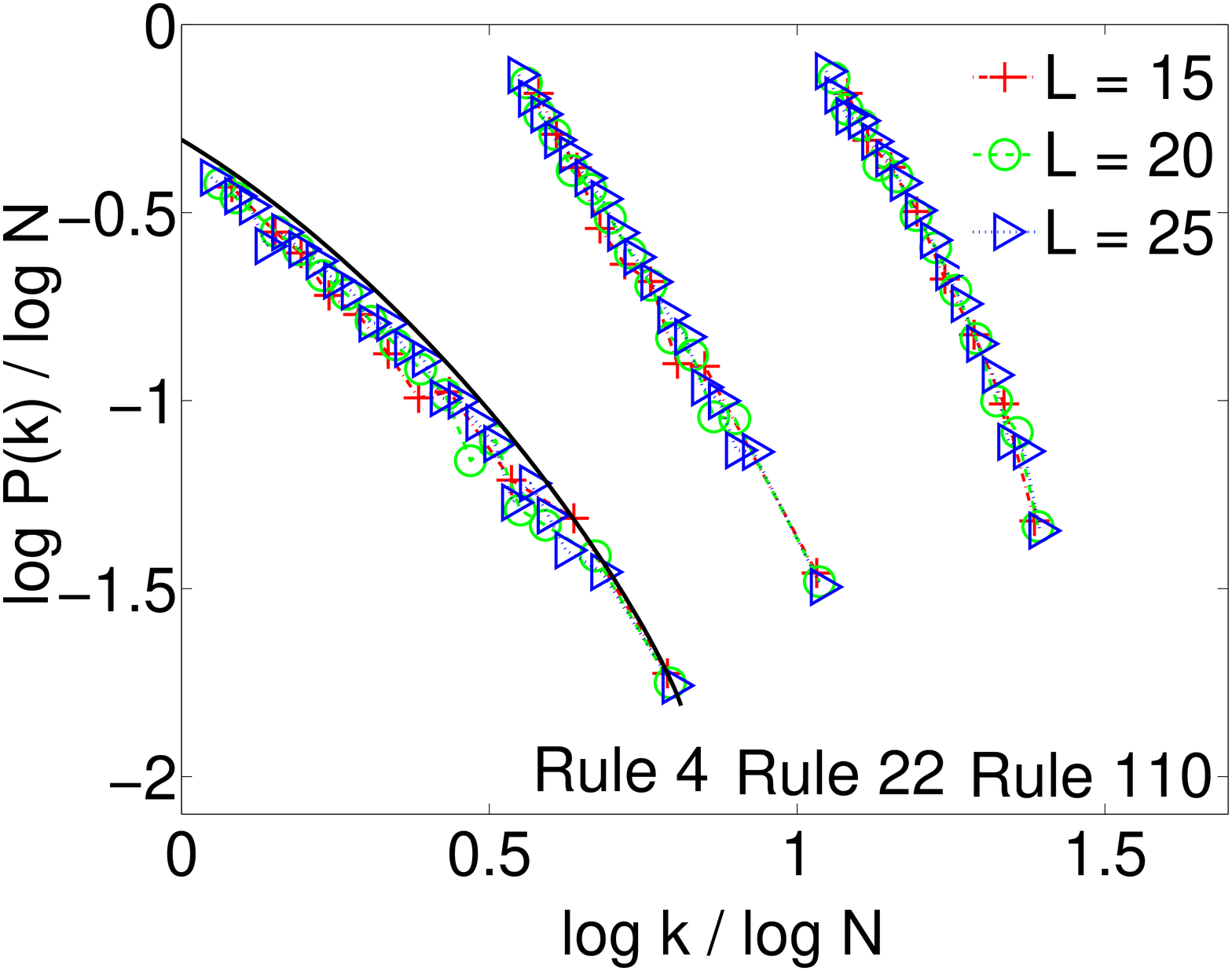}
\caption{(color on-line) In-degree distribution functions collapsed for simple and complex
  rules using a multiscaling ansatz for rules 4, 22, and 110 for different system 
  sizes.  The black solid line is Eq.~(\ref{eq: y of x}).
  The distributions were shifted by 0.5 units each for clarity.
  Rule 22 is consistent with a power law $P(k) \sim k^{-\gamma}$ with
  $\gamma \approx 2.8$. }
\label{collapse}
\end{center}
\end{figure}

Another interesting property of rules giving large
hubs is the scaling 
of the in-degree distribution function $P(k)$ with system size.  
Fig.~\ref{collapse} shows the data collapse for Rules 4, 22, and 110
found using the multiscaling ansatz~\cite{Lise-Pac}
\begin{equation}
   \log P(k) = \log N \; f(\log k / \log N)\;.
\end{equation}
It is significantly better than the usual finite size power-law scaling, 
which would produce straight lines in 
Fig.~\ref{collapse}. To decide whether the apparent curvature is a
finite size effect or a true indication of multiscaling is in general 
not easy, but it can be done analytically for the simple CA, Rule 4.

For Rule 4, the sequence $\dots 11\ldots$ has no pre-image and the
sequence $\ldots 1\ldots$ has the unique pre-image $\ldots 010\ldots$.
${\bf T}^{(1)}$ can then be written as ${\bf e}_2^{\top} {\bf e}_3$,
where ${\bf e}_2 = (0,1,0,0)$ and ${\bf e}_3 = (0,0,1,0)$.
The in-degree of any state ${\bf S}$ can then be expressed as 
\begin{equation}
   \prod_i w(m_i) = \prod_i {\bf e}_3 \, [{\bf T}^{(0)}]^{m_i} \, {\bf e}_2^{\top},
\end{equation}
where 
\begin{equation}
{\bf T}^{(0)} = \left( \begin{array}{cccc}
1 & 1 & 0 & 0 \\
0 & 0 & 0 & 1 \\
1 & 1 & 0 & 0 \\
0 & 0 & 1 & 1
\end{array}
\right)\;,
\end{equation}
the product runs over all `1's in ${\bf S}$, and $m_i$ is the number of 
`0's following the $i$th `1' in ${\bf S}$. 

For large $m$ we have $w(m) \approx a \lambda^m$, where $\lambda =
1.75488$ is the largest eigenvalue of ${\bf T}^{(0)}$ and
$a = 0.234487 = ({\bf e}_2 \cdot {\bf u}) ({\bf v}\cdot {\bf e}_3)$,
${\bf u}$ and ${\bf v}$ being the right and left eigenvectors
corresponding to $\lambda$, normalized to ${\bf v} \cdot {\bf u}=1$.
Any state containing $n$ isolated 1s will then have an in-degree $k_n 
\approx \prod_{i=1}^n a\lambda^{m_i} = a^n \lambda^{L-n}$. Although
occurrences of small $m_i$ can only be neglected for $n\ll L$,
we find empirically that this formula reflects the qualitative behavior
for larger $n$. To find
$P(k)$, we now need to find $\Omega(n)$, the number of states with $n$
isolated 1s and no pairs `11', and transform its dependence on $n$
into a dependence on $k$.  On a periodic lattice of length $L$, one
has $\Omega(n) = C(L-n,n)+C(L-n-1,n-1)$, where $C(\ell,m)=
\ell!/[m!(\ell-m)!]$. To obtain an approximation for $P(k)$ we first
invert the above relation between $k_n$ and $n$, giving $ n(k) = (L
\ln\lambda - \ln k)/[\ln (\lambda/a)]$, then use $P(k) \approx
2^{-L}\Omega(n)|dn/dk|$.

The scaling ansatz shown in Fig.~\ref{collapse} is indeed recovered 
by this approximation. To see this, we define
\begin{equation}
   x \equiv \ln k/\ln N, \qquad y \equiv \ln P(k)/\ln N~.
\end{equation}
With this choice, we have
\begin{equation}
\label{eq: n of x}
  \frac n L = \frac{\ln\lambda - x\ln 2}{\ln (\lambda/a)}
\end{equation}
Neglecting a term $\ln\ln(\lambda/a)/(L\ln 2)$ in $y$,
using Stirling's formula, and taking the large $L$ limit of
$\ln[\Omega(n)]/L$ for fixed $x$, we get
\begin{equation}
\label{eq: y of x}
   y \approx -1 - x + \log_2 \left[\frac{(1-\epsilon)^{1-\epsilon}}
         {\epsilon^\epsilon(1-2\epsilon)^{1-2\epsilon}}\right].
\end{equation}
Here $\epsilon \equiv n/L$  is a function of $x$ through
Eq.~(\ref{eq: n of x}).  Eq.~(\ref{eq: y of x}) is shown as a solid
line in Fig.~\ref{collapse}. The curvature of this line clearly
indicates that no substantial range exists over which the
distribution is a power law.  Using a Fourier method based on
recursion relations, we have also determined numerically the exact
distribution for $L=10,000$.  It shows slightly enhanced curvature for
small $x$, and is for large $x$ in excellent agreement with the above
approximation. Hence we conclude that the apparent curvature seen in Fig.
4 for Rules 4 and 110, is not simply a finite size effect but more likely
an indication of multiscaling. 

\begin{figure}[htbp]
\begin{center}
\includegraphics*[width=2.1in, height=3.3in, angle=270]{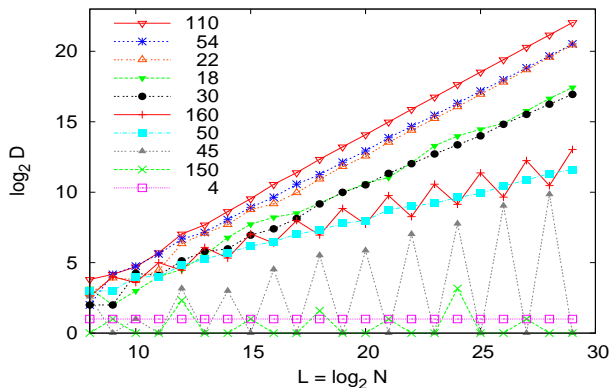}
\caption{(color on-line) The path diversity $\cal D$ of different CA. 
   Straight lines indicate scaling ${\cal D} \sim N^{\delta}$. Fitted 
   values of $\delta$ are $0.88, 0.88, 0.85, 0.75$ and 0.72 for 
   rules 110, 22, 54, 18 and 30, respectively.}
\label{diversity}
\end{center}
\end{figure}

As Fig.~1 indicates, rule 4 does not exhibit heterogeneity beyond the 
single node level. All transients have length 1, all attractors are 
simple fixed points, and the only `complex' aspect is the broad
in-degree distribution of the latter. Clearly, the in-degree
distribution of any set of nodes, being a strictly local construct, 
cannot by itself distinguish complex CA from trivial ones. To make this 
distinction we introduce a new quantity, the {\it path diversity} 
${\cal D}$.
It measures fluctuations in the set of different paths connecting the 
GoE states to attractors. If one projects all attractor states into
a single node, then the state space network of a CA becomes a rooted
tree. Roughly, ${\cal D}$ counts the number of non-equivalent choices 
encountered by following each path from an attractor 
(root) to a GoE state (leaf). Path diversity bears resemblances to 
tree diversity~\cite{Huberman} 
and topological depth~\cite{Badii}.

We first define the path diversity ${\cal D}$ for each transient node: 
A GoE state has diversity one; a
transient state with a single pre-image has the same diversity as its
unique pre-image; and the diversity of a node with more than one incoming
link is the sum of all \emph{distinct} diversities of its pre-images
plus one. Thus, if a node has e.g. in-degree 5 and three of its
pre-images have diversity 2, one has diversity 6, and the last has
diversity 17, then that node's diversity is $2+6+17+1=26$. Finally,
the path diversity ${\cal D}$ of the entire CA is computed by joining all
attractor states (in all disconnected components, if there are several) into one single ``meta-state", and applying the above scheme to
the meta-state.  

In Fig.~\ref{diversity}, $\cal D$ is shown for several CA. 
It is sensitive to aspects of the network's structure that are different from  the node degree distribution. As a result, it clearly separates rules 4 (where 
${\cal D} =2$ for all $L$) and 150 (where ${\cal D}$ seems to remain 
bounded) from other CA.
The most interesting rules are those which show clear scaling
${\cal D} \sim N^\delta$ with $\delta$ close to 1, e.g. ${\delta} = 
0.88\pm 0.01$ (rules 22, 110), $0.85\pm 0.01$ (rule 54), $0.75\pm 0.04$ 
(rule 18), and $0.72\pm 0.03$ (rule 30). 
For the system sizes studied, rules 50 and 160 appear to show scaling
with smaller $\delta$, though both would be classified as simple
(classes I or II) by Wolfram.  An analytical calculation of upper
bounds on transient lengths shows, however, that ${\cal D}$ actually
grows slower than any positive power of $N$ for rule 160, and we
expect the same to hold for rule 50 given its similar numerics.  The
oscillations seen in Fig.~\ref{diversity} for some rules are due to
the global constraint imposed by periodic boundary conditions.  In the
most extreme case, rule 45 has nontrivial transients for odd $L$ but
none at all for even $L$.

In general, we neither expect that a single observable can reliably
distinguish ``complex" from other behavior, nor that a few discrete
classes can do justice to the multifarious ways in which a system can 
display complexity. This is supported by our analysis. Neither 
the scaling of the hub sizes with system size (or the scaling of the 
in-degree distribution) nor the scaling of the path diversity can by 
itself distinguish between ``simple" 
(Wolfram classes I and II) and ``complex" (class IV and some class III)
dynamics. Combined together, they fare already much better. 
We note also that
within the class of CA for which $\cal D$ scales, 
those with largest ${\delta}$ (CA 110
and 22) are also those which have been considered most complex
by previous authors \cite{Cook,Peter1986JStatPhys}.

In summary, we have studied networks formed by states
of elementary 1-d CA on finite lattices. We find that some of them 
exhibit non-trivial scaling with system size, and this scaling behavior
can be computed analytically 
in certain cases. 
Taken together, the the statistics of degree distribution (a local property) and path diversity (a global property)
give a good indication of dynamical complexity. 
Including additional measures for network heterogeneity at intermediate
scales could further refine this approach.



Part of this work was done at the Perimeter Institute for Theoretical Physics. 
J.E.S.S.~and B.S.~were supported by NSF Grant PHY-0417372 and the Institute for 
Biocomplexity and Informatics at the University of Calgary. W.N. was supported 
by NSF Grant CHE-0313618.

\bibliography{References}
\end{document}